\documentclass[preprint,letterpaper,preprintnumbers,aps,superscriptaddress,nofootinbib,tightenlines,floatfix]{revtex4}

\usepackage{amsmath,amssymb}
\usepackage{graphicx}
\usepackage{bm}
\usepackage{comment}

\newcommand{\beq}{\begin{equation}}
\newcommand{\eeq}{\end{equation}}
\newcommand{\bea}{\begin{eqnarray}}
\newcommand{\eea}{\end{eqnarray}}

\def\OMIT#1{{}}

\newcount\hour \newcount\hourminute \newcount\minute 
\hour=\time \divide \hour by 60
\hourminute=\hour \multiply \hourminute by 60
\minute=\time \advance \minute by -\hourminute
\newcommand{\mydate}{\ \today \ - \number\hour :\number\minute}

\begin{document}



\preprint{\vbox{ \hbox{JLAB-THY-10-1121} \hbox{NT@UW-10-01} }}

\title{A method to study complex systems of mesons in Lattice QCD}

\author{William Detmold} \affiliation{Department of Physics, College
  of William and Mary, Williamsburg, VA 23187-8795.}
\affiliation{Jefferson Laboratory, 12000 Jefferson Avenue, Newport
  News, VA 23606.}  \author{Martin J.~Savage} \affiliation{Department
  of Physics, University of Washington, Seattle, WA 98195-1560.}

\date{\mydate}

\begin{abstract}
  \noindent
  Finite density systems can be explored with Lattice QCD through the
  calculation of multi-hadron correlation functions.  Recently,
  systems with up to 12 $\pi^+$'s or $K^+$'s have been studied to
  determine the $3$-$\pi^+$ and $3$-$K^+$ interactions, and the
  corresponding chemical potentials have been determined as a function
  of density.  We derive recursion relations between correlation
  functions that allow this work to be extended to systems of
  arbitrary numbers of mesons and to systems containing many different
  types of mesons, such as $\pi^+$'s, $K^+$'s, $\overline{D}^0$'s and
  $B^+$'s. These relations allow for the study of finite-density
  systems in arbitrary volumes, and for the study of high-density
  systems.
\end{abstract}
\maketitle

%
%
\section{Introduction }
\label{sec:Intro}

\noindent 
An important goal of Lattice QCD (LQCD) is to calculate, 
with  quantifiable uncertainties, the properties and
interactions
of systems comprised of multiple hadrons directly from QCD.  
The last few years have
seen the first calculations of three-baryon systems in
QCD~\cite{Beane:2009gs} ($\Xi\Xi n$, and the triton or $^3$He), a
four-baryon system in quenched QCD~\cite{Yamazaki:2009ua} (the
$\alpha$-particle), and the
three-$\pi^\pm$~\cite{Beane:2007es,Detmold:2008fn} and
three-$K^\pm$~\cite{Detmold:2008yn} interactions from the calculation
of systems involving up to twelve
$\pi^\pm$'s~\cite{Beane:2007es,Detmold:2008fn} and
$K^\pm$'s~\cite{Detmold:2008yn} respectively.  While all of these
calculations were at unphysical values of the light quark masses due
to the limited computational resources, they represent a significant
step forward in a QCD-based understanding of the complex hadronic
systems that dominate nature.

The study of multi-meson systems comprised of one or more species will
provide important insights into the structure of dense forms of matter
that may arise in astrophysical settings.  Further, they will provide
insight into the phase structure of QCD, and strongly interacting
many-body systems in general.  Finite density systems of mesons have
been studied in LQCD using an appropriate chemical potential
\cite{Kogut:2004zg,Sinclair:2006zm,deForcrand:2007uz}.  However, as
shown in Refs.~\cite{Beane:2007es, Detmold:2008fn, Detmold:2008yn},
one can also study these systems as a function of density and chemical
potential by explicitly considering LQCD correlation functions with
increasing numbers of mesons.  For instance, the isospin chemical
potential has been determined as a function of isospin density from
systems of $\pi^+$'s \cite{Detmold:2008fn} by measuring the
ground-state energies of different numbers of mesons in a fixed
volume, and forming discrete differences, e.g. $\mu_I\sim dE/dn\sim
(E_{n+j}-E_n)/j$.

Lattice QCD calculations of systems involving multiple hadrons, such
as nuclei or systems of multiple mesons, necessarily involve large
numbers of contractions between quark field operators which naively
grow as the product of the factorial of the number of each flavor of
quark present in the system.  For instance, a simple interpolating
field for the proton is comprised of two up-quarks and one down-quark,
and therefore the number of independent contractions required in the
computation of the proton correlation function is $N_{p} = (2!)(1!) =
2$. The proton-proton correlation function requires $N_{pp} = (4!)(2!)
= 48$, the triton ($pnn$) correlation function (or, equivalently in
the isospin limit, $^3{\rm He}$) requires $N_{pnn} = (4!)(5!) = 2880$,
and the $\alpha$-particle ($ppnn$) requires $N_{ppnn} = (6!)(6!) =
518400$.  In the first calculation of three-baryon
systems~\cite{Beane:2009gs}, the $\Xi\Xi n$ and the triton, the number
of measurements of the correlation function that could be made was
limited, not by the number of gauge-field configurations or quark
propagators that could be computed, but by the number of contractions
that could be performed with the available computational resources
(even after identifying identical and vanishing contributions).  The
same limitation exists for the calculation of systems involving large
numbers of mesons.  The actual number of contractions required for
such systems can be substantially reduced by exploiting the symmetry
of the contractions~\cite{Beane:2009gs, Yamazaki:2009ua, Beane:2007es,
  Detmold:2008fn, Detmold:2008yn} (identifying redundant
contributions), or by using different sources (e.g. using only the
upper two components of the quark field
operators~\cite{Yamazaki:2009ua}).  However, even with these
simplifications, the number of contractions does not scale
polynomially with the number of hadrons to large systems, and the
calculation of contractions remains a significant roadblock to the
exploration of multi-hadron systems with LQCD.

In this work, we develop recursion relations among contractions that
allow for  the calculation of correlation functions corresponding to
systems with arbitrary numbers of mesons.\footnote{\label{ft} In this
  work, we limit our discussion to mesonic systems that do not involve
  creation and annihilation of the same flavor of quark field at the
  same Euclidean time. We also only consider pseudoscalar mesons for
  simplicity, however, there are no conceptual difficulties in
  including other types of mesons. To be specific, we focus on mesons
  with the the quantum numbers of $\overline{q}\gamma_5 u$ with $q\ne
  u$.} 
The correlation function of the $({\cal N}+1)$-meson system is related 
to that of the ${\cal N}$-meson system by a small
number of matrix and scalar multiplications using the recursion relations.  
The recursion makes use
of the fact that many of the contractions required for the 
$({\cal N}+1)$-meson system have already been calculated in the construction
of the ${\cal N}$-meson system.  
The simplest recursion relations for
a single species of meson are developed in Section
\ref{sec:onequarkprop}, and the generalizations to two species and to many
species are presented in Sections \ref{sec:twoquarkprop} and
\ref{sec:mquarkprop}. As the repeated use of a quark propagator from a
single source limits the number of mesons in the system to be ${\cal
  N}\le N_c\ N_s =12$ (where $N_c$ and $N_s$ are the number of colors
and spinor components, respectively), we present recursion relations
for systems arising from two sources in Section \ref{sec:twosrcs} and
from multiple sources in Section \ref{sec:msrcs}. These two extensions
of the original recursions are finally combined into a recursion
relation that allows for systems with arbitrary numbers of mesons of arbitrary
species (see footnote \ref{ft}) to be computed from
propagators from many different sources. This is presented in Section
\ref{sec:msrcskspecies}.  The recursive approach offers a significant
speedup for intermediate-sized systems and allows for the investigation of
larger systems that are otherwise impractical. Section
\ref{sec:discussion} is a concluding discussion of such computational
aspects and summarizes the broader perspective of this approach.

\section{Single Species Multi-Meson Systems from One Source}
\label{sec:onequarkprop}

\noindent
Let us begin by considering multi-pion systems that are composed of
$n$-$\pi^+$'s for which the correlation functions are produced from a
single light-quark propagator.  As there are $N_s=4$ Dirac indices and
$N_c=3$ color indices associated with each quark field (on a given
lattice site), there are $N_s\times N_c=12$ independent components in
each quark-field and hence a single light-quark propagator can be used
to generate systems containing up to $12$ $\pi^+$'s.  To calculate
systems with $n>12$, additional distinct light-quark propagators must
be calculated, as discussed below.  A correlation function for a system of $n<12$
$\pi^+$'s has the form
\begin{eqnarray}
  C_{n\pi^+}(t) & = &  
  \left\langle\ \left(\ \sum_{\bf x}\ \pi^+({\bf x},t)\ \right)^n\ 
  \left( \phantom{\sum_{\bf x}}\hskip -0.2in
    \pi^-({\bf  0},0)\ \right)^n\ \right\rangle
  \ \ \ ,
  \label{eq:npip}
\end{eqnarray}
where the operator $\pi^+ ({\bf x},t)$ denotes a quark-level operator
$\pi^+ ({\bf x},t) = \overline{d}({\bf x},t) \ \gamma_5\ u ({\bf
  x},t)$.  
Naively, there are $N_{\overline{d}}! N_u! = (n!)^2$
independent contractions that 
contribute to 
this correlation function, which for the $n=12$ system corresponds to
a total number of $\sim 2.3\times 10^{17}$.  By the symmetry of the
correlation function, with all propagators originating from a single
source, all of the contractions of either the up- or down-quark fields
are the same, leaving only $n!$ contractions to be evaluated, which
for $n=12$ is $\sim 4.8\times 10^8$.  However, considering how the
contractions can be grouped by permutations, there are far fewer
independent contractions that must be performed.

As the sources and the sinks of each meson are identical, a twelve component
Grassmann variable, $\eta$, can be introduced in order to write the
correlation function in eq.~(\ref{eq:npip}) as
\begin{eqnarray}
  C_{n\pi^+}(t) & = & n!\  
  \langle\ \left(\ \overline{\eta}_i\ A_{ij} (t)\ \eta_j\ \right)^n\ \rangle
  \ \ ,\ \ 
  A_{ij} (t) \ = \ \sum_{\bf x}\ \left[S({\bf x},t;{\bf
      0},0)\right]_{ik}\ \left[S^\dagger ({\bf x},t;{\bf
    0},0)\right]_{kj}\ 
  \ \ \ \ ,
  \label{eq:npipGRASSMANN}
\end{eqnarray}
where $S({\bf x},t;{\bf 0},0)$ is the light-quark propagator from the
source located at $({\bf 0},0)$ to the sink at $({\bf x},t)$ and we
have used the relation $S({\bf x},t;{\bf y},t^\prime)=\gamma_5
S^\dagger({\bf y},t^\prime;{\bf x},t)\gamma_5$. $A(t)$ is a
time-dependent $12\times12$ matrix and the indices in
eq.~(\ref{eq:npipGRASSMANN}) are combined spinor-color indices running
over $i=1,\ldots,12$. Repeated spinor-color indices imply
summation. From the anti-commuting nature of of the $\eta$, it
follows that
\begin{eqnarray}
  C_{n\pi^+}(t) & = & (-)^n\ {n!\over (N-n)!}\ 
  \epsilon^{a_1 ...a_{N-n}\alpha_1 ... \alpha_n } \  
  \epsilon_{a_1 ...a_{N-n}\beta_1 ... \beta_n }  \ 
  \left[ \ A(t)\ \right]_{\alpha_1 \beta_1}  ...
  \left[ \ A(t)\ \right]_{\alpha_n \beta_n} 
  \nonumber\\
 \label{eq:npiDET}
\end{eqnarray}
where $N=N_s\times N_c$, and the indices $a_i$, $\alpha_i$, and
$\beta_i$ are summed.  In the case
of interest, $N=12$, but the relations that we derive are true for
arbitrary values of $N$. 
An important building block for the contractions $C_{n\pi^+}$ is a
partly contracted object $R_n$ whose spinor-color trace is (up
to an irrelevant combinatorial factor) equivalent to the
contraction. Formally this is defined via the functional relation
\begin{equation}
  \label{eq:1}
  \left[R_n\right]_{ij}=\overline{u}_i(0)d_k(0)\frac{\delta}
{\delta d_k(0)}\frac{\delta}{
\delta\overline{u}_j(0) }C_{n\pi^+}
\end{equation}
 The correlation functions in
eq.~(\ref{eq:npiDET}) can be related to sums of traces over $A(t) $ as
\begin{eqnarray} {\rm det}\left[\ 1\ +\ \lambda \ A\ \right]
  & = & {1\over N!}\ \sum_j^N\ ^NC_j\ \lambda^j\ 
  \epsilon^{a_1 ...a_{N-j}\alpha_1 ... \alpha_j } \  
  \epsilon_{a_1 ...a_{N-j}\beta_1 ... \beta_j }  \ 
  \left[ \ A(t)\ \right]_{\alpha_1\beta_1}  ...
  \left[ \ A(t)\ \right]_{\alpha_j\beta_j} 
  \nonumber\\
  & = & 
  {\rm exp}\left( {\rm Tr}\left[\ \log\left[ \ 1\ +\ \lambda \ A\ \right]
    \right]\right)
  \ =\ 
  {\rm exp}\left( {\rm Tr}\left[\ \sum_{p=1}\ {(-)^{p-1}\over p}\ \lambda^p\
      A^p\ \right]
  \right)
  \nonumber\\
  & = & 
  1\ +\ 
  \lambda\ \langle A\rangle\ +\ {\lambda^2\over 2!}\left( \langle A\rangle^2 -
    \langle A^2\rangle\right)
  \ +\ {\lambda^3\over 3!}\left(\langle A\rangle^3 - 3 \langle A^2\rangle
    \langle A\rangle + 2 \langle A^3\rangle\right)
  \nonumber\\
  &  & \ +\ ... 
  \nonumber\\
  & = & 
  \sum_{j}^N\ {1\over j!}\ \lambda^j\ \langle\ R_j\ \rangle
  \ =\ 
  \sum_{j}^N\ (-)^j\ 
  \left( {1\over j!}\right)^2 \ \lambda^j\ C_{j\pi^+}
  \ \ \ \ ,
  \label{eq:npiREL}
\end{eqnarray}
with the explicit expressions for systems with $n\leq 13$ given in the
Appendix of Ref.~\cite{Detmold:2008fn}.  In the last line $\langle
R_j\rangle\equiv {\rm Tr}[R_n]$ is the Dirac and color trace. As the
l.h.s. of eq.~(\ref{eq:npiREL}) is an order-N polynomial in $\lambda$,
the $\langle\ R_j\ \rangle=0$ (and hence $C_{j\pi^+}=0$) $\forall\ j\
>\ N$.  For $n=12$, there are approximately $80$ independent terms
that must be summed (resulting from the partition of $12$
objects)~\cite{Beane:2007es}, requiring approximately $10^3$
calculations, which is significantly smaller than the naive number of
$\sim 2.3\times 10^{17}$ and the improved number of $\sim 4.8\times
10^8$.  One sees that large coefficients appear in the expansion of
$\langle R_n\rangle$ for large values of $n$, leading to significant
cancellations among terms, and the need to use high precision
arithmetic libraries in the numerical calculation of such correlation
functions.  For large $n$, the number of terms that must be evaluated
behaves as ${1\over 2 \sqrt{2 n} \pi }\ e^{\pi\sqrt{2n/3}}$
\cite{HardyRamanujan}, which scales poorly to systems involving a
large number of $\pi^+$'s.

\subsection{Ascending Recursion Relations}
\label{sec:onequarkpropASC}

\noindent
The objects $R_n$ in eq.~(\ref{eq:npiREL}) are $N\times N$ matrices,
and their trace is proportional to the contraction associated with the
$n$-pion system.  The matrices themselves correspond to the
contractions in the $n$-pion system with one up-quark and one
anti-up-quark remaining uncontracted.  Therefore, the contraction
associated with the $(n+1)$-pion system can be found by contracting
$A$ with $R_n$ in all possible ways.  There are two independent
contractions of $A$ and $R_n$, and their coefficients can be
determined by requiring that the $\langle R_n\rangle$ reproduce the
multi-$\pi^+$ contractions given in Ref.~\cite{Detmold:2008fn}.  It is
straightforward to show that the object $R_{n+1}$ associated with the
$(n+1)$-$\pi^+$ system is related to that of the $n$-$\pi^+$ system
through
\begin{eqnarray}
  R_{n+1} & = & \langle\ R_n\ \rangle\ A\ -\ n\ R_n\ A \, .
  \label{eq:npipRECURSION}
\end{eqnarray}
In order for the recursion relation in eq.~(\ref{eq:npipRECURSION}) to
be useful for LQCD calculations, a starting point (starting
contraction) must be identified.  An obvious starting point is the
contraction associated with the single-$\pi^+$ system, $n=1$, for
which $R_1=A$, and $\langle R_1\rangle = \langle A\rangle$.  This
can be used as the starting point of ascending recursion relations
that determine $\langle R_{n+1}\rangle$ from $R_{n}$.  On the other
hand, a less obvious starting point is that $R_{N+1}=0$, induced by
the Pauli-principle, which will yield descending recursion relations
from which $R_{n-1}$ can be determined from $R_n$.

The initial condition for the ascending recursion relation is (beyond
$\langle\ R_0\ \rangle=1$)
\begin{eqnarray}
  R_1 & = & A
  \ \ ,\ \ 
  \langle R_1\rangle =
  \langle A\rangle
  \ \ ,
  \label{eq:pipR1}
\end{eqnarray}
The correlation
function for the $2$-$\pi^+$ system is
\begin{eqnarray}
  R_2 & = & \langle\ R_1\ \rangle A \ -\ R_1\ A
  \ =\ \langle\ A\ \rangle\ A \ -\ A^2
  \nonumber\\
  \langle \ R_2\ \rangle & = &
  \ \langle\ A\ \rangle^2 \ -\ \langle\ A^2\ \rangle
  \ \ ,
  \label{eq:pipR2}
\end{eqnarray}
which agrees with the result in Ref.~\cite{Detmold:2008fn}.  The correlation
function for the $3$-$\pi^+$ system is
\begin{eqnarray}
  R_3 & = & \langle\ R_2\ \rangle\  A \ -\ 2\ R_2\ A
  \ =\  
  \langle\ A\ \rangle^2 \ A
  \ -\ \langle\ A^2\ \rangle\ A
  \ -\ 2\ \langle\ A\ \rangle\ A^2 
  \ +\ 2\ A^3\ 
  \nonumber\\
  \langle\ R_3 \ \rangle 
  & = & 
  \langle\ A\ \rangle^3
  \ -\ 3\ \langle\ A^2\ \rangle\ \  \langle\ A\ \rangle\
  \ -\ 2\ \langle\ A^3\ \rangle\ 
  \ \ ,
  \label{eq:pipR3}
\end{eqnarray}
also in agreement with Ref.~\cite{Detmold:2008fn}.  Repeated
application of the recursion relation recovers all of 
the contractions given
explicitly in Ref.~\cite{Detmold:2008fn}.

\subsection{Descending Recursion Relations}
\label{sec:onequarkpropDeSC}

\noindent
The ascending recursion relation enables a sequential calculation of
the correlation functions for systems containing $n$-$\pi$'s for $n\le
N$ from a single light-quark propagator.  For $n>N$ the correlations
functions all vanish due to the Pauli-principle, which is implemented
by eq.~(\ref{eq:npiREL}).  As $R_{N+k}=0$ for $k > 0$ for an arbitrary
matrix $A$, it is obvious from the recursion relation,
eq.~(\ref{eq:npipRECURSION}), that $R_N\propto I_N$, where $I_N$ is
the $N\times N$ identity matrix.  It then follows from
eq.~(\ref{eq:npiREL}) that
\begin{eqnarray}
  R_N & = & (N-1)!\ {\rm det}\left(\ A\ \right)\ I_N\,,
  \nonumber\\
  \langle\ R_N\ \rangle & = & N!\ {\rm det}\left(\ A\ \right)
  \ \ \ .
  \label{eq:Rndef}
\end{eqnarray}
The fact that $R_N\propto I_N$ and $ R_{N+1}=0$ allows one to
construct descending recursion relations by working with ``holes'' in
the ``closed-shell'' of $R_N$.  Multiplying the recursion relation in
eq.~(\ref{eq:npipRECURSION}) by $A^{-1}$ on the right yields
\begin{eqnarray}
  R_{n+1}\ A^{-1} & = & \langle\ R_n\ \rangle\ I_N\ -\ n\ R_n
  \nonumber\\
  \langle\ R_{n+1}\ A^{-1}\ \rangle & = & \left(\ N - n\ \right)\langle\ R_n\ \rangle\ 
  \ \ \ ,
  \label{eq:Rnp1}
\end{eqnarray}
from which it follows that
\begin{eqnarray}
  R_{n-1} & = & {1\over n-1}\ 
  \left[\ {1\over N+1-n}\ \langle\ R_{n}\ A^{-1}\ \rangle \ I_N\ -\ R_{n}\
    A^{-1}\ \right]
  \ \ \ ,
  \label{eq:Rnm1}
\end{eqnarray}
and therefore provides a descending recursion relation where
$A^{-1}$ is interpreted as a $\pi^+$-hole (while $A$ is
interpreted as a $\pi^+$).  
Applying this recursion relation to the
result in eq.~(\ref{eq:Rndef}) produces
\begin{eqnarray}
  R_{N-1} & = & (N-2)!\ {\rm det}\left(\ A\ \right)\ 
  \left[\ 
    \langle\ A^{-1}\ \rangle\ I_N
    \ - \ 
    A^{-1}\ \right]
  \nonumber\\
  \langle\ R_{N-1}\ \rangle 
  & = & (N-1)!\ {\rm det}\left(\ A\ \right)\ 
  \langle\ A^{-1}\ \rangle\ 
  \ \ \ ,
  \label{eq:RNm1}
\end{eqnarray}
and further application of the recursion relation
to the result in eq.~(\ref{eq:RNm1}) produces
\begin{eqnarray}
  R_{N-2} & = &{(N-3)!\over 2}\ {\rm det}\left(\ A\ \right)\ 
  \left[ 
    \langle\ A^{-1}\ \rangle^2 I_N
    - 
    \langle\ \left(A^{-1}\right)^2\ \rangle I_N
    - 2\ \langle\ A^{-1}\ \rangle\ A^{-1}
    + 2\  \left(A^{-1}\right)^2 
  \right]
  \nonumber\\
  \langle\ R_{N-2}\ \rangle & = &{(N-2)!\over 2}\ {\rm det}\left(\ A\ \right)\ 
  \left[\ 
    \langle\ A^{-1}\ \rangle^2\ - \ \langle\ \left(A^{-1}\right)^2\ \rangle\ \right]
  \ \ \ .
  \label{eq:RNm2}
\end{eqnarray}
It is interesting to note that the $\langle\ R_{N-k}\ \rangle$ have
the same form in terms of the $A^{-1}$ as the $\langle\ R_{k}\
\rangle$ do in terms of $A$ (modulo factors of ${\rm det}\left(\
  A\ \right)\ $ and numerical factors depending upon $k, N$).  This
observation makes it obvious that
\begin{eqnarray}
  \langle\ R_{N-k}\ \rangle & = & {(N-k)!\over k!}\ {\rm det}\left(\ A\ \right)\ 
  \langle\ {\cal R}_{k}(A^{-1})\ \rangle
  \ \ \ ,
  \label{eq:RNmk}
\end{eqnarray}
where the recursion ${\cal R}_{k}(w)$ is defined by
\begin{eqnarray} {\cal R}_{n+1}(w) & = & \langle\ {\cal R}_n(w)\
  \rangle\ w\ -\ n\ {\cal R}_n\ w \ \ \ .
  \label{eq:npipRECURSIONgen}
\end{eqnarray}

\section{Two Species Multi-Meson Systems from One Source}
\label{sec:twoquarkprop}

\noindent
The recursion relations that allow for the computation of correlation
functions for systems composed of $(n+1)$-$\pi^+$'s from systems
composed of $n$-$\pi^+$ can be extended to construct the correlation
functions composed of both $\pi^+$'s and $K^+$'s.  A correlation
function for a system composed of $n_\pi$ $\pi^+$'s and $n_K$ $K^+$'s
is
\begin{eqnarray}
  C_{\{ n_\pi\pi^+\ ,\ n_K K^+ \}}(t) & = &  
  \Bigg\langle\ 
  \left(\ \sum_{\bf x}\ \pi^+({\bf x},t)\ \right)^{n_\pi} \ 
  \left(\ \sum_{\bf x}\ K^+({\bf x},t)\ \right)^{n_K} \ 
  \nonumber\\
  && \qquad \qquad \qquad \qquad \qquad 
  \left( \phantom{\sum_{\bf x}}\hskip -0.2in
    \pi^-({\bf  0},0)\ \right)^{n_\pi}\ 
  \left( \phantom{\sum_{\bf x}}\hskip -0.2in
    K^-({\bf  0},0)\ \right)^{n_K}\ 
  \Bigg\rangle
  \ \ \ ,
  \label{eq:npipnKK}
\end{eqnarray}
where the operator $K^+ ({\bf x},t)$ denotes a quark-level operator
$K^+ ({\bf x},t) = \overline{s}({\bf x},t) \ \gamma_5\ u ({\bf x},t)$.
After contracting the quark field operators, the correlation function
can be written as
\begin{eqnarray}
  C_{\{ n_\pi\pi^+\ ,\ n_K K^+ \}}(t) & = &  n_\pi !\ n_K !\   
  \langle\ 
  \left(\ \overline{\eta}\ A (t)\ \eta\ \right)^{n_\pi}\ 
  \left(\ \overline{\eta}\ \kappa (t)\ \eta\ \right)^{n_K}\ 
  \rangle
  \nonumber\\ 
  \kappa (t) & = & \sum_{\bf x}\ S({\bf x},t;{\bf 0},0)\ S_s^\dagger ({\bf x},t;{\bf
    0},0)\ 
  \ \ \ \ ,
  \label{eq:npipGRASSMANNpiK}
\end{eqnarray}
where $S_s({\bf x},t;{\bf 0},0)$ is the strange quark propagator from
the source located at $({\bf 0},0)$ to the sink at $({\bf x},t)$.  The
factor of $n_\pi !\ n_K !$ that appears (instead of the $n!$ in
eq.~(\ref{eq:npipGRASSMANN})) corresponds to the number of ways of
contracting both the anti-strange and anti-down light-quark field
operators.  Setting $n=n_\pi+n_K$ in eq.~(\ref{eq:npipGRASSMANN}) and
eq.~(\ref{eq:npiDET}), making the replacement $\lambda A\rightarrow\
\lambda A + \beta\kappa$ and identifying terms that are of the same
order in $\lambda^j\ \beta^{n-j}$, we find that
\begin{eqnarray}
  C_{\{ n_\pi\pi^+\ ,\ n_K K^+ \}}(t) & = &   (-)^{n_\pi+n_K}\ 
  { n_\pi !\ n_K !\   \over (N-n_\pi-n_K)!}\ 
  \epsilon^{a_1 ...a_{N-n}\alpha_1 ... \alpha_{n_\pi+n_K} } \  
  \epsilon_{a_1 ...a_{N-n}\beta_1 ... \beta_{n_\pi+n_K} }  \ 
  \nonumber\\
  && \qquad  \qquad  \qquad  \qquad   
  \left[ \ A(t)\ \right]_{\alpha_1}^{\beta_1}  ...
  \left[ \ A(t)\ \right]_{\alpha_{n_\pi}}^{\beta_{n_\pi}} 
  \left[ \ \kappa(t)\ \right]_{\alpha_{n_\pi+1}}^{\beta_{n_\pi+1}}  ...
  \left[ \ \kappa(t)\ \right]_{\alpha_{n_\pi+n_K}}^{\beta_{n_\pi+n_K}} 
  \nonumber\\
  & = & 
  (-)^{n_\pi+n_K}\  
  {n_\pi !\ n_K !\over ^{n_\pi+n_K}C_{n_\pi}}
  \ \langle\ R_{\{n_\pi , n_K\} }\ \rangle
  \ \ \ \ ,
  \label{eq:npiDETpik}
\end{eqnarray}
where $R_{ \{ n_\pi , n_K\} }$ is the generalization of $R_n$ to the
two-species system.  By construction, we are restricted to systems
with $n_\pi+n_K\le N$ for propagators from single sinks.  As the
recursion relation in eq.~(\ref{eq:npipRECURSION}) is satisfied under
the replacement $\lambda A\rightarrow\ \lambda A + \beta\kappa$, it is
clear that the $R_{ \{ n_\pi , n_K\} }$ satisfy a recursion relation
\begin{eqnarray}
  R_{ \{ n_\pi  , n_K\} }
  & = & 
  \langle\ R_{ \{ n_\pi -1 , n_K\} }\ \rangle\ A
  \ -\ 
  (n_\pi+n_K-1)\ R_{ \{ n_\pi -1 , n_K\} }\ A
  \nonumber\\
  & + &
  \langle\ R_{ \{ n_\pi   , n_K - 1\} }\ \rangle\ \kappa
  \ -\ (n_\pi+n_K-1)\ R_{ \{ n_\pi   , n_K - 1\} }\ \kappa
  \ \ \ \ .
  \label{eq:piKRECURSION}
\end{eqnarray}
%
The boundary
conditions for the ascending recursion relations are
\begin{eqnarray}
  R_{ \{ 1 , 0\} } & = & A
  \ \ ,\ \ 
  R_{ \{ 0 , 1\} } \ = \ \kappa
  \ \ ,\ \ 
  \langle\ R_{ \{ 0 , 0\} } \ \rangle\ =\ 1
  \ \ \ \ ,
  \label{eq:piKBCs}
\end{eqnarray}
and  $R_{ \{ p , -j\} }=0$ and$R_{ \{ -j , p\} }=0$ $\forall\ j >
0$ and $\forall\ p$.

The descending recursion relations are a little less obvious.  Unlike
the case of $N$ $\pi^+$'s for which there is a single system with
$n_\pi=N$, the mixed $\pi^+$-$K^+$ systems has a set of systems with
$n_\pi+n_K=N$.  It remains the case that $R_{j,N-j+1}=0$ $\forall\ j$,
and further, the single species results provide
\begin{eqnarray}
  R_{\{N,0\}} & = & (N-1)!\ {\rm det}\left(A\right)\ I_N
  \ \ ,\ \ 
  R_{\{0,N\}} \ = \ (N-1)!\ {\rm det}\left(\kappa\right)\ I_N
  \ \ \ \ .
  \label{eq:piKNbc}
\end{eqnarray}
Using the replacement $\lambda A\rightarrow\
\lambda A \left( 1 + {\beta\over\lambda}A^{-1}\kappa\right)$ in
eq.~(\ref{eq:npipRECURSION}) we see that
\begin{eqnarray}
  R_{\{N-j,\ j\} } & = & { (N-1)!\over j!}\  {\rm det}\left(A\right)\ 
  \langle\ {\cal R}_j(A^{-1}\kappa)\ \rangle \ I_N 
  \nonumber\\
  & = & 
  { (N-1)!\over (N-j)! }\  {\rm det}\left(\kappa\right)\ 
  \langle\ {\cal R}_{N-j}(\kappa^{-1}A)\ \rangle \ I_N 
  \ \ \ \ ,
  \label{eq:piKNj}
\end{eqnarray}
which allows for the contractions of the systems with $n_\pi+n_K=N$ to
be related to each other.

To reduce the total number of mesons in the system to $n_\pi+n_K<N$
requires use of eq.~(\ref{eq:piKRECURSION}), which 
can be written as
\begin{eqnarray}
  && R_{\{N-p-j,\ j\}} \ =\  
  {1\over N-p} \ 
  \left[\ 
    {1\over p}\ 
    \left( \ 
      \langle R_{\{N-p-j+1,j\}}\ A^{-1} \rangle
      \ -\ 
      \langle R_{\{N-p-j+1,j-1\}}\rangle\ 
      \langle \kappa A^{-1} \rangle
    \right.
  \right.
  \nonumber\\
  && \left.
    \left. 
      \qquad  
      +  
      (N-p)\langle R_{\{N-p-j+1,j-1\}}\ \kappa A^{-1}\rangle\ 
    \right)
    \ I_N 
  \right.
  \\
  && \left.\qquad  
    - R_{\{N-p-j+1,j\}}\ A^{-1}
    + 
    \langle R_{\{N-p-j+1,j-1\}}\rangle\ 
    \kappa A^{-1}
    -
    (N-p)R_{\{N-p-j+1,j-1\}}\kappa A^{-1}
    \phantom{1\over p}
  \right]
  \nonumber
  \ \ \ \ ,
  \label{eq:piKdescending}
\end{eqnarray}
from which the correlation function with $(N-p-j)$ $\pi^+$'s and $j$
$K^+$'s can be determined from the correlation functions with
$(N-p-j+1)$ $\pi^+$'s and $j$ $K^+$'s, and $(N-p-j+1)$ $\pi^+$'s and $(j-1)$
$K^+$'s.  For instance, as we have expressions for the system with
$n_\pi$ $\pi^+$'s and $0$ $K^+$'s, and also for the system with $(N-1)$
$\pi^+$'s and $1$ $K^+$'s, the relation in
eq.~(\ref{eq:piKdescending}) can be used to obtain the correlation
function for the system with $(N-2)$ $\pi^+$'s and $1$ $K^+$'s,
\begin{eqnarray}
  &&R_{\{N-2,1\}} \ = \ 
  {1\over N-1}\ \left[\ 
  \right.
  \nonumber\\
  &&\left.\qquad\qquad
    \left(\ 
      \langle\ R_{\{N-1,1\}}\ A^{-1}\ \rangle
      \ -\ 
      \langle\ R_{\{N-1,0\}}\ \rangle\ \langle\ \kappa A^{-1}\ \rangle
      +(N-1)\ \langle\ R_{\{N-1,0\}}\ \kappa A^{-1}\ \rangle
    \right)\ I_N\ 
  \right.
  \nonumber\\
  &&\left.\qquad\qquad
    \ -\ R_{\{N-1,1\}}\ A^{-1}\ 
    \ +\ 
    \langle\ R_{\{N-1,0\}}\ \rangle\ \kappa A^{-1}\ 
    \ -\ (N-1)\ R_{\{N-1,0\}}\ \kappa A^{-1}\ 
    \ 
  \right]
  \nonumber\\
  &&  \phantom{R_{\{N-2,1\}} }
  = (N-2)!\ {\rm det}\left(A\right)
  \left[\ 
    \left(\ 
      \langle\ A^{-1}\ \rangle \langle\ \kappa A^{-1}\ \rangle 
      \ -\ 
      \langle\ A^{-1}\kappa A^{-1}\ \rangle 
    \right)\ I_N
  \right.
  \nonumber\\
  &&\left.\qquad\qquad\qquad\qquad\qquad\qquad\qquad
    \ +\ A^{-1}\kappa A^{-1}\ -\  \langle\ \kappa A^{-1}\ \rangle A^{-1}\
  \right]\,,
  \nonumber\\
  \nonumber\\
  && \langle\ R_{\{N-2,1\}} \ \rangle 
  \ = \ 
  \langle\ R_{\{N-1,1\}}\ A^{-1}\ \rangle
  \ -\ 
  \langle\ R_{\{N-1,0\}}\ \rangle\ \langle\ \kappa A^{-1}\ \rangle
  +(N-1)\ \langle\ R_{\{N-1,0\}}\ \kappa A^{-1}\ \rangle
  \nonumber\\
  && \phantom{\langle\ R_{\{N-2,1\}} \ \rangle }
  \ = \ 
  (N-1)!\ {\rm det} \left(A\right)\ 
  \left[\ 
    \langle\ A^{-1}\ \rangle\ 
    \langle\ \kappa A^{-1}\ \rangle\ 
    \ - \ 
    \langle\ A^{-1}\kappa A^{-1}\ \rangle\ 
    \ \right]
  \ \ \ \ .
  \label{eq:piKdescending21}
\end{eqnarray}
Once this is known, it can be combined with the correlation function
for $(N-2)$ $\pi^+$'s and $2$ $K^+$'s, to produce that for $(N-3)$
$\pi^+$'s and $2$ $K^+$'s, and so forth, determining the correlation
functions for all systems with $n_\pi+n_K=N-1$. This process can then
be repeated to produce the correlation functions for all systems with
$n_\pi+n_K=N-2$, $n_\pi+n_K=N-3$, and so forth.  The fact that we have
calculated the correlation functions for purely $\pi^+$-systems,
purely $K^+$ systems and mixed systems with a total of $N$ $\pi^+$'s
and $K^+$'s, allows for the correlation functions for all systems with
$n_\pi+n_K\le N$ to be determined from descending recursion relations.

\section{m- Species Multi-Meson Systems from One Source}
\label{sec:mquarkprop}

\noindent
It is now possible to generalize the discussions of the previous
sections, and arrive at the correlation functions for systems
comprised of mesons of more than one species, generated with a
single light-quark propagator, and multiple different light, strange
or heavy quark propagators.  This allows for the discussions of
systems comprised of, for instance, $n_\pi$ $\pi^+$'s, $n_K$ $K^+$'s,
$n_D$ $\overline{D}^0$'s, 
and $n_B$ $B^+$'s.  A correlation function for a system
composed of $n_1$ mesons of type ${\cal A}_1$, $n_2$ mesons of type
${\cal A}_2$, ..., $n_m$ mesons of type ${\cal A}_m$, is of the form
\begin{eqnarray}
  C_{\{ n_1 {\cal A}_1\ ,\ ...\ , n_m {\cal A}_m\}}(t) & = &  
  \Bigg\langle\ 
  \left(\ \sum_{\bf x}\ {\cal A}_1({\bf x},t)\ \right)^{n_1} \ ...\  
  \left(\ \sum_{\bf x}\ {\cal A}_m({\bf x},t)\ \right)^{n_m} \ 
  \nonumber\\
  && \qquad \qquad \qquad \qquad
  \left( \phantom{\sum_{\bf x}}\hskip -0.2in
    {\cal A}_1^\dagger({\bf  0},0)\ \right)^{n_1}\ ...\ 
  \left( \phantom{\sum_{\bf x}}\hskip -0.2in
    {\cal A}_m^\dagger({\bf  0},0)\ \right)^{n_m}\ 
  \Bigg\rangle
  \ \ \ ,
  \label{eq:mm}
\end{eqnarray}
where the operator ${\cal A}_m ({\bf x},t)$ denotes a quark-level
operator ${\cal A}_m ({\bf x},t) = \overline{q}_m({\bf x},t) \
\gamma_5\ u ({\bf x},t)$.  After contracting the quark field
operators, this can be written as
\begin{eqnarray}
  C_{\{ n_1 {\cal A}_1\ ,\ ...\ , n_m {\cal A}_m\}}(t) & = &  
  n_1 !\ ...\ n_m !\   
  \langle\ 
  \left(\ \overline{\eta}\ A_1 (t)\ \eta\ \right)^{n_1}\ ...\ 
  \left(\ \overline{\eta}\ A_m (t)\ \eta\ \right)^{n_m}\ 
  \rangle\,,
  \nonumber\\ 
  A_j (t) & = & \sum_{\bf x}\ S({\bf x},t;{\bf 0},0)\ S_j^\dagger ({\bf x},t;{\bf
    0},0)\ 
  \ \ \ \ ,
  \label{eq:npipGRASSMANNmm}
\end{eqnarray}
where $S_j({\bf x},t;{\bf 0},0)$ is the propagator of the $j^{\rm th}$
quark flavor from the source located at $({\bf 0},0)$ to the sink at
$({\bf x},t)$.
Writing the total number of mesons in the system as $ \sum_i\ n_i =
{\cal N}$, we have that
\begin{eqnarray}
  && C_{\{ n_1 {\cal A}_1\ ,\ ...\ , n_m {\cal A}_m\}}(t) \ =\  
  (-)^{\cal N}\ 
  { \prod_i\  n_i !\   \over (N-{\cal N})!}\ 
  \epsilon^{a_1 ...a_{N-{\cal N}}\alpha_1 ... \alpha_{\cal N} } \  
  \epsilon_{a_1 ...a_{N-{\cal N}}\beta_1 ... \beta_{\cal N} }  \ 
  \nonumber\\
  && \ \   
  \left[ \ A_1(t)\ \right]_{\alpha_1}^{\beta_1}  ...
  \left[ \ A_1(t)\ \right]_{\alpha_{n_1}}^{\beta_{n_1} }
  \left[ \ A_2(t)\ \right]_{\alpha_{n_1+1}}^{\beta_{n_1+1}}  ...
  \left[ \ A_2(t)\ \right]_{\alpha_{n_1+n_2}}^{\beta_{n_1+n_2}}... 
  \left[ \ A_m(t)\ \right]_{\alpha_{{\cal N}-n_m}}^{\beta_{{\cal N}-n_m}}  ...
  \left[ \ A_m(t)\ \right]_{\alpha_{{\cal N}}}^{\beta_{{\cal N}}} 
  \nonumber\\
  &  &
  \qquad\qquad
  \ =\  
  (-)^{\cal N}\ 
  { \left( \ \prod_i\  n_i ! \ \right)^2\   \over {\cal N}!}\ 
  \ \langle\ R_{\{n_1 , ... , n_m\} }\ \rangle
  \ \ \ \ ,
  \label{eq:npiDETmm}
\end{eqnarray}
where $R_{\{n_1 , ... , n_m\} }$ is the generalization of $R_n$ to the
$m$-species system.  The $R_{\{n_1 , ... , n_m\} }$ satisfy a set of 
recursion relations such as
\begin{eqnarray}
  R_{\{n_1+1, n_2 , ... , n_m\} }
  & = & 
  \langle\ R_{\{n_1, n_2 , ... , n_m\} }\  \rangle\ A_1\ 
  \ -\ {\cal N}\ R_{\{n_1, n_2 , ... , n_m\} }\ A_1\ 
  \nonumber\\
  & + & \
  \langle\ R_{\{n_1+1, n_2-1 , ... , n_m\} }\  \rangle\ A_2\ 
  \ -\ {\cal N}\ R_{\{n_1+1, n_2-1 , ... , n_m\} }\ A_2\ 
  \ +\ ...
  \nonumber\\
  & + & \
  \langle\ R_{\{n_1+1, n_2 , ... , n_k-1 , ... , n_m\} }\  \rangle\ A_k\ 
  \ -\ {\cal N}\ R_{\{n_1+1, n_2 , ... , n_k-1 , ... , n_m\} }\ A_k\ +\ ...
  \nonumber\\
  & + & \
  \langle\ R_{\{n_1+1, n_2 , ... , n_m-1\} }\  \rangle\ A_m\ 
  \ -\ {\cal N}\ R_{\{n_1+1, n_2 , ... , n_m-1\} }\ A_m\ 
  \ \ \ \ ,
  \label{eq:mmREC}
\end{eqnarray}
which are an obvious generalization of eq.~(\ref{eq:piKRECURSION}).
These can be written more compactly as
\begin{eqnarray}
  R_{\{ {\bf n}+1_k  \} }
  & = & 
  \sum_{j=1}^m\ 
  \langle\ 
  R_{\{ {\bf n}+1_k-1_j  \} }
  \ \rangle
  \ A_j
  \ -\ 
  {\cal N}\ R_{\{ {\bf n}+1_k-1_j  \} }\  A_j
  \ \ \ ,
  \label{eq:mmRECcompact}
\end{eqnarray}
where $\{ {\bf n}\}=\{n_1,n_2,...,n_m\}$, and $\{ {\bf
  n}+1_k\}=\{n_1,n_2,...,n_k+1, ...,n_m\}$.

As may be guessed from the complexity of the descending recursion
relations in the two-species case, the descending recursion relations
in the m-species case are quite unpleasant, and we do not present them.

\section{Single Species Multi-Meson Systems Beyond $n=12$}
\label{sec:multisrcs}

\noindent
Calculations using a single source for quark propagators are limited
to systems involving $n\le 12$ $\pi^+$'s.  Systems comprised
of $n>12$ $\pi^+$'s can be studied by computing light-quark
propagators produced from more than one source.  For instance, systems
with $n\le 24$ $\pi^+$'s can be studied by working with light-quark
propagators produced from two different sources and, more generally,
systems with $n\le 12\ p$ $\pi^+$'s can be studied by working with
light-quark propagators produced from $p$ different sources.  
It is then obvious that to study
a system of $240$ $\pi^+$'s will require the calculation of
light-quark propagators from $20$ different sources.

\subsection{Single Species Multi-Meson Systems from Two Sources}
\label{sec:twosrcs}

Instead of considering propagators from only a single source point at
$({\bf 0},0)$, we can consider propagators from two source points at
$({\bf y}_1,0)$ and $({\bf y}_2,0)$.  A correlation function for a
system of $\overline{n}=n_1+n_2$ $\pi^+$'s with $n_1$ emanating from
$({\bf y}_1,0)$ and $n_2$ emanating from $({\bf y}_2,0)$ is
\begin{eqnarray}
  C_{( n_1\pi^+_1\ ,\ n_2 \pi^+_2 )}(t) & = &  
  \Bigg\langle\ 
  \left(\ \sum_{\bf x}\ \pi^+({\bf x},t)\ \right)^{n_1+n_2} \ 
  \left( \phantom{\sum_{\bf x}}\hskip -0.2in
    \pi^-({\bf  y_1},0)\ \right)^{n_1}\ 
  \left( \phantom{\sum_{\bf x}}\hskip -0.2in
    \pi^-({\bf  y_2},0)\ \right)^{n_2}\ 
 \Bigg\rangle
  \ .
  \label{eq:pipi}
\end{eqnarray}
After contracting the quark field operators, the correlator can be
written as
\begin{eqnarray}
  C_{( n_1\pi^+_1\ ,\ n_2 \pi^+_2 )}(t) & = &  
  \overline{n}!\   
  \langle 
  \left(\ 
    \overline{\eta} 
P_1(t)
    \eta\ \right)^{n_1} 
  \left(\ 
    \overline{\eta} 
P_2(t)
    \eta\ \right)^{n_2} 
  \rangle
  \ ,
  \label{eq:npipGRASSMANNpipi}
\end{eqnarray}
where $\eta$ is now a $24$-component Grassmann variable, corresponding
to the 12-components of the up-quark field at position $({\bf y}_1,0)$
and the 12-components of the up-quark field at position $({\bf
  y}_2,0)$. The $24\times24$ matrices 
\begin{eqnarray}
  P_1 & = & 
  \left(\begin{array}{c|c}
      A_{11}(t)&A_{12}(t)\\
      \hline
      0&0
    \end{array}
  \right)
  \ \ ,\ \ 
  P_2\ =\ 
  \left(\begin{array}{c|c}
      0&0\\
      \hline
      A_{21}(t)&A_{22}(t)
    \end{array}
  \right)
  \ \ \ \ ,
  \label{eq:Pdef}
\end{eqnarray}
are constructed from the $12\times 12$ matrices
\begin{eqnarray}
  &&A_{ij}  =  \sum_{\bf x}\ S({\bf x},t;{\bf y}_j,0)\ S^\dagger ({\bf x},t;{\bf y}_i,0)\ 
\,,
  \label{eq:pipiblocks}
\end{eqnarray}
with $i,j$ denoting the propagator source locations. It is straightforward to show
that
\begin{eqnarray}
  C_{( n_1\pi^+_1\ ,\ n_2 \pi^+_2 )}(t) & = &  
  (-)^{\overline{n}}\ 
  {\overline{n}! \over (\overline{N}-\overline{n})!}\ 
  \epsilon^{a_1 ...a_{\overline{N}-\overline{n}}\alpha_1 ... \alpha_{\overline{n}} } \  
  \epsilon_{a_1 ...a_{\overline{N}-\overline{n}}\beta_1 ... \beta_{\overline{n}} }  \ 
  \nonumber\\
  && \qquad  \qquad  \qquad  \qquad   
  \left[ \ P_1(t)\ \right]_{\alpha_1}^{\beta_1}  ...
  \left[ \ P_1(t)\ \right]_{\alpha_{n_1}}^{\beta_{n_1}} 
  \left[ \ P_2(t)\ \right]_{\alpha_{n_1+1}}^{\beta_{n_1+1}}  ...
  \left[ \ P_2(t)\ \right]_{\alpha_{\overline{n}}}^{\beta_{\overline{n}}} 
  \nonumber\\
  & = & 
  (-)^{\overline{n}}\ 
  {\overline{n}! \over ^{\overline{n}}C_{n_1}} \ 
  \langle\ Q_{(n_1 , n_2) }\ \rangle
  \ \ \ \ ,
  \label{eq:pipieps}
\end{eqnarray}
where we have defined $\overline{N} = 2 N$ ($=24$).
It is obvious that if either $n_1>N$ ($>12$) or $n_2>N$, then the
correlation function vanishes.  The $Q_{(n_1 , n_2) }$ are
$\overline{N}\times\overline{N}$ matrices (that are time-dependent),
and satisfy the recursion relation
\begin{eqnarray}
  Q_{( n_1+1 , n_2 )} & = & 
  \langle\ Q_{(n_1 , n_2 )}\ \rangle\ P_1\ 
  - \ (n_1+n_2)\ Q_{(n_1 , n_2 )}\ \ P_1
  \nonumber\\
  & + &
  \langle\ Q_{(n_1 +1 , n_2 - 1 )}\ \rangle\ P_2\ 
  - \ (n_1+n_2)\ Q_{(n_1 +1 , n_2 -1  )}\ \ P_2
  \ \ \ \ ,
  \label{eq:Qrecur}
\end{eqnarray}
and a similar relation for $Q_{(n_1,n_2+1)}$.
The boundary condition for the recursion relation in
eq.~(\ref{eq:Qrecur}) is
\begin{eqnarray}
  Q_{( 1 , 0 )}  & = & P_1
  \ \ ,\ \ 
  Q_{( 0 , 1 )}  \ = \ P_2
 \ \ ,\ \ 
  \langle\ Q_{( 0 , 0 )}\ \rangle  \ = \ 1
  \ \ \ \ ,
  \label{eq:QrecurBC}
\end{eqnarray}
with $Q_{(j,k)}=0$ if either $j<0$ or $k<0$.
This recursion relation is somewhat less obvious than those that
describe systems with a single light-quark propagator, and it is worth
demonstrating its implementation.  For the $n_1+n_2=2$ systems, the
recursion relation gives
\begin{eqnarray}
  Q_{( 2 , 0 )} & = & 
  \langle\ Q_{(1 , 0 )}\ \rangle\ P_1\ 
  - \ \ Q_{(1 , 0 )}\ \ P_1
  \nonumber\\
  & = & 
  \left\langle\ 
  \left(\begin{array}{c|c}
      A_{11}&A_{12}\\
      \hline
      0&0
    \end{array}
  \right)
  \ \right\rangle 
  \left(\begin{array}{c|c}
      A_{11}&A_{12}\\
      \hline
      0&0
    \end{array}
  \right)
  \ -\ 
  \left(\begin{array}{c|c}
      A_{11}&A_{12}\\
      \hline
      0&0
    \end{array}
  \right)
  \left(\begin{array}{c|c}
      A_{11}&A_{12}\\
      \hline
      0&0
    \end{array}
  \right)
  \nonumber\\
  & = & 
  \left(\begin{array}{c|c}
      \langle\ A_{11}\ \rangle A_{11} \ -\ A_{11}^2
      & \langle\ A_{11}\ \rangle A_{12}\ -\ A_{11}\  A_{12}
      \\
      \hline
      0&0
    \end{array}
  \right)\,,
  \nonumber\\
  \langle\ Q_{( 2 , 0 )}\ \rangle
  & = & 
  \langle\ A_{11}\ \rangle^2 \ -\ \langle\ A_{11}^2\ \rangle
  \ \ \ \ .
  \label{eq:Q20}
\end{eqnarray}
\begin{eqnarray}
  Q_{( 0 , 2 )} & = & 
  \langle\ Q_{(0 , 1 )}\ \rangle\ P_2\ 
  - \ \ Q_{(0 , 1 )}\ \ P_2
  \nonumber\\
  & = & 
  \left\langle\ 
  \left(\begin{array}{c|c}
      0&0\\
      \hline
      A_{21}&A_{22}
    \end{array}
  \right)
  \ \right\rangle
  \left(\begin{array}{c|c}
      0&0\\
      \hline
      A_{21}&A_{22}
    \end{array}
  \right)
  \ -\ 
  \left(\begin{array}{c|c}
      0&0\\
      \hline
      A_{21}&A_{22}
    \end{array}
  \right)
  \left(\begin{array}{c|c}
      0&0\\
      \hline
      A_{21}&A_{22}
    \end{array}
  \right)
  \nonumber\\
  & = & 
  \left(\begin{array}{c|c}
      0&0\\
      \hline
      \langle\ A_{22}\ \rangle\ A_{21}\ -\ A_{22}\ A_{21}
      &\langle\ A_{22}\ \rangle\ A_{22}\ -\ A_{22}^2
    \end{array}
  \right)\,,
  \nonumber\\
  \langle\ Q_{( 0 , 2 )}\ \rangle
  & = & 
  \langle\ A_{22}\ \rangle^2 \ -\ \langle\ A_{22}^2\ \rangle
  \ \ \ \ .
  \label{eq:Q02}
\end{eqnarray}
\begin{eqnarray}
  Q_{( 1 , 1 )} & = & 
  \langle\ Q_{(0 , 1 )}\ \rangle\ P_1\ 
  - \ \ Q_{(0 , 1 )}\ \ P_1
  \ +\ 
  \langle\ Q_{(1 , 0 )}\ \rangle\ P_2\ 
  - \ \ Q_{(1 , 0 )}\ \ P_2
  \nonumber\\
  & = & 
  \left\langle\ 
  \left(\begin{array}{c|c}
      0&0\\
      \hline
      A_{21}&A_{22}
    \end{array}
  \right)
  \ \right\rangle\ 
  \left(\begin{array}{c|c}
      A_{11}&A_{12}\\
      \hline
      0&0
    \end{array}
  \right)
  \ -\ 
  \left(\begin{array}{c|c}
      0&0\\
      \hline
      A_{21}&A_{22}
    \end{array}
  \right)
  \ 
  \left(\begin{array}{c|c}
      A_{11}&A_{12}\\
      \hline
      0&0
    \end{array}
  \right)
  \nonumber\\
  & + & 
  \left\langle\ 
  \left(\begin{array}{c|c}
      A_{11}&A_{12}\\
      \hline
      0&0
    \end{array}
  \right)
  \ \right\rangle\ 
  \left(\begin{array}{c|c}
      0&0\\
      \hline
      A_{21}&A_{22}
    \end{array}
  \right)
  \ -\ 
  \left(\begin{array}{c|c}
      A_{11}&A_{12}\\
      \hline
      0&0
    \end{array}
  \right)
  \  
  \left(\begin{array}{c|c}
      0&0\\
      \hline
      A_{21}&A_{22}
    \end{array}
  \right)
  \nonumber\\
  & = & 
  \left(
    \begin{array}{c|c}
      \langle\ A_{22}\ \rangle\ A_{11}\ -\ A_{12}A_{21}
      &
      \langle\ A_{22}\ \rangle\ A_{12}\ -\ A_{12}A_{22}
      \\
      \hline
      \langle\ A_{11}\ \rangle\ A_{21}\ -\ A_{21}A_{11}
      &
      \langle\ A_{11}\ \rangle\ A_{22}\ -\ A_{21}A_{12}
    \end{array}
  \right)\,,
  \nonumber\\
  \langle\ Q_{( 1 , 1 )}\ \rangle 
  & = & 
  2\ \left[\ \langle\ A_{22}\ \rangle\ \langle\ A_{11}\ \rangle\ 
    - \ \langle \ A_{12}A_{21}\ \rangle\ 
  \right]
  \ \ \ \ .
  \label{eq:Q11}
\end{eqnarray}
The result for $\langle\ Q_{( 1 , 1 )}\ \rangle $ obtained in
eq.~(\ref{eq:Q11}) exhibits the expected result when the second source
is set to be identical to the first, reproducing the single source
result multiplied by the combinatoric factor of $^2C_1$.  Repeated
application of the recursion relation generates all contractions
possible from the two sources.  In the case of three mesons, we find
that
\begin{eqnarray}
  &&{1\over ^3C_2}\ \langle Q_{( 2 , 1 )}\rangle  =  
  \left[ 
    \langle A_{11} \rangle^2 \langle A_{22} \rangle
    - 
    \langle A_{11}^2 \rangle \langle A_{22} \rangle
    + 
    2 \langle A_{11}A_{12} A_{21} \rangle
    - 
    2 \langle A_{12}A_{21} \rangle \langle  A_{11} \rangle
  \right]
  \ ,
  \label{eq:Qn3}
\end{eqnarray}
which recovers the single-source result when ${\bf y}_2 ={\bf y}_1$.

\subsection{Single Species Multi-Meson Systems from m Sources}
\label{sec:msrcs}

The extension of the two-source result in eq.~(\ref{eq:Qrecur}), which
provided a way to explore systems comprised of up to $n\le 24$
$\pi^+$'s, to systems generated with $m$-sources can be achieved with
a similar construction.  The correlation function for a system of
$\overline{n}=\sum_i\ n_i$ $\pi^+$'s is
\begin{eqnarray}
  C_{( n_1\pi^+_1\ , ... , \ n_m \pi^+_m )}(t) & = &  
  \langle\ 
  \left(\ \sum_{\bf x}\ \pi^+({\bf x},t)\ \right)^{\overline{n}}  
  \left( \phantom{\sum_{\bf x}}\hskip -0.2in
    \pi^-({\bf  y_1},0)\ \right)^{n_1} ...  
  \left( \phantom{\sum_{\bf x}}\hskip -0.2in
    \pi^-({\bf  y_m},0)\ \right)^{n_m}\ 
  \rangle
  \ ,
  \label{eq:mpi}
\end{eqnarray}
which is equal to
\begin{eqnarray}
  && 
  C_{( n_1\pi^+_1\ , ... , \ n_m \pi^+_m )}(t) \ =\ 
  \overline{n}!\   
  \langle\ 
  \left(\ 
    \overline{\eta}\ 
    P_1
    \ \eta\ \right)^{n_1}\ ...\ 
  \left(\ 
    \overline{\eta}\ P_m
    \ \eta\ \right)^{n_m}\ 
  \rangle\,,
\end{eqnarray}
where the $\eta$ are now $m\times N$ component Grassmann variables and the
\begin{eqnarray}
  P_k = 
  \left(
    \begin{array}{c|c|c|c}
      0&0 & ... & 0\\
      \hline
      \vdots & \vdots & ... & \vdots \\
      \hline
      A_{k1}(t)&A_{k2}(t)&\ \  ... \ \ & A_{km}(t)\\
      \hline
      0&0& ... & 0\\
      \hline
      \vdots & \vdots & ... & \vdots \\
      \hline
      0&0& ... & 0
    \end{array}
  \right)
  \ ,
  \label{eq:GRASSMANNmpi}
\end{eqnarray}
with the $A_{ij}(t)$ defined in eq.~(\ref{eq:pipiblocks}).  This can
be expressed as
\begin{eqnarray}
  && C_{( n_1\pi^+_1\ , ... , \ n_m \pi^+_m )}(t) \ =\  
  (-)^{\overline{n}}\ 
  {\overline{n}! \over (\overline{N}-\overline{n})!}\ 
  \epsilon^{a_1 ...a_{\overline{N}-\overline{n}}\alpha_1
    ... \alpha_{n_1}\alpha_{n_1+1}...\alpha_{\overline{n}} } \ 
  \epsilon_{a_1 ...a_{\overline{N}-\overline{n}}\beta_1   ... 
    \beta_{n_1}\beta_{n_1+1}...\beta_{\overline{n}} }  \ 
  \nonumber\\
  && \qquad  \qquad  \qquad  \qquad   
  \left[ \ P_1(t)\ \right]_{\alpha_1}^{\beta_1}  ...
  \left[ \ P_1(t)\ \right]_{\alpha_{n_1}}^{\beta_{n_1}} 
  \left[ \ P_2(t)\ \right]_{\alpha_{n_1+1}}^{\beta_{n_1+1}}  ...
  \left[ \ P_2(t)\ \right]_{\alpha_{n_1+n_2}}^{\beta_{n_1+n_2}}  ...
  \left[ \ P_m(t)\ \right]_{\alpha_{\overline{n}}}^{\beta_{\overline{n}}} 
  \nonumber\\
  & & 
  \phantom{C_{( n_1\pi^+_1\ , ... , \ n_m \pi^+_m )}(t)}
  \ = \ (-)^{\overline{n}}\ 
  \left(\ \prod_i\ n_i! \ \right)\ 
  \langle\ Q_{(n_1 , n_2 , ... , n_m) }\ \rangle
  \ \ \ \ ,
  \label{eq:mpieps}
\end{eqnarray}
where $\overline{N} = m\ N$.  The $Q_{(n_1 , n_2 , ... , n_m)}$
satisfy the recursion relation
\begin{eqnarray}
  Q_{(n_1+1 , n_2 , ... , n_m)} & = & 
  \langle\ Q_{(n_1 , n_2 , ... , n_m)}\ \rangle\ P_1
  \ -\ 
  \overline{n}\ Q_{(n_1 , n_2 , ... , n_m)}\ P_1
  \nonumber\\
  & ... & +\ 
  \langle\ Q_{(n_1+1 , n_2 , ... n_k-1, ... ,  n_m)}\ \rangle\ P_k
  \ -\ 
  \overline{n}\ Q_{(n_1+1 , n_2 , ... n_k-1 , ... , n_m)}\ P_k
  \nonumber\\
  & ... & +\ 
  \langle\ Q_{(n_1+1 , n_2 , ... , n_m-1)}\ \rangle\ P_m
  \ -\ 
  \overline{n}\ Q_{(n_1+1 , n_2 , ... , n_m-1)}\ P_m
  \ \ \ \ ,
  \label{eq:mpiRECUR}
\end{eqnarray}
which can be written in a more concise way as
\begin{eqnarray}
  Q_{( {\bf n} +1_k )} & = & 
  \sum_{i=1}^m\ 
  \langle\ Q_{( {\bf n} +1_k - 1_i  )} \ \rangle P_i
  \ -\ 
  \overline{n}\ Q_{( {\bf n} +1_k - 1_i  )} \  P_i
  \ \ \ \ ,
  \label{eq:mpiRECURcompact}
\end{eqnarray}
where $( {\bf n})=(n_1,n_2,...,n_m)$, and $( {\bf
  n}+1_k)=(n_1,n_2,...,n_k+1, ...,n_m)$, {\it etc}.  The recursion
relation in eq.~(\ref{eq:mpiRECUR}) allows for the calculation of
systems involving large numbers of $\pi^+$'s.  As an example, the
application of the recursion relation to the contraction for the
$3$-$\pi^+$ systems resulting from 3 different sources reproduces the
correct result of
\begin{eqnarray} {1!1!1! \over 3!} Q_{(1 , 1 , 1)} & = & \left[\
    \langle\ A_{11}\ \rangle \langle\ A_{22}\ \rangle \langle\
    A_{33}\ \rangle \ \right.
  \nonumber\\
  && \left.  \ -\ \langle\ A_{12}A_{21}\ \rangle \langle\
    A_{33}\ \rangle \ -\ \langle\ A_{13}A_{31}\ \rangle \langle\
    A_{22}\ \rangle \ -\ \langle\ A_{23}A_{32}\ \rangle \langle\
    A_{11}\ \rangle \ \right.  \nonumber\\ && \left.  \ +\ \langle\
    A_{12}A_{23}A_{31}\ \rangle \ +\ \langle\
    A_{13}A_{32}A_{21}\ \rangle \ \right] \ \ \ \ ,
  \label{eq:threesrc}
\end{eqnarray}
and recovers the single-source result when ${\bf y}_3={\bf y}_2={\bf y}_1$.

\section{k Species Multi-Meson Systems from m Sources}
\label{sec:msrcskspecies}

\noindent
In this section, we generalize the results of the previous sections to
the correlation functions of systems composed of arbitrary numbers of
species of mesons with the quantum numbers of $\overline{q}_i\gamma_5
u$ for $q_i\ne u$ (for instance, systems comprised of $\pi^+$'s,
$K^+$'s, $\overline{D}^0$'s, $B^+$'s) resulting from an arbitrary
number of light-quark sources.  A correlation function for a system
composed of $n_{ij}$ mesons of the $i^{\rm th}$ species from the
$j^{\rm th}$ source at $({\bf y}_j,0)$, where $0\le i\le k$ and $0\le
j\le m$, is of the form
\begin{eqnarray}
  && C_{\bf n}(t) \ = \  
  \Bigg\langle\ 
  \left(\ \sum_{\bf x}\ {\cal A}_1 ({\bf x},t)\ \right)^{{\cal N}_1} \ 
  ...
  \left(\ \sum_{\bf x}\ {\cal A}
    _k ({\bf x},t)\ \right)^{{\cal N}_k} \
  \nonumber\\
  &&
  \left( \phantom{\sum_{\bf x}}\hskip -0.22in
    {\cal A}_1^\dagger({\bf  y_1},0) \right)^{n_{11}} ...\  
  \left( \phantom{\sum_{\bf x}}\hskip -0.22in
    {\cal A}_1^\dagger({\bf  y_m},0) \right)^{n_{1m}} ...\  
  \left( \phantom{\sum_{\bf x}}\hskip -0.22in
    {\cal A}_k^\dagger({\bf  y_1},0) \right)^{n_{k1}} ...\  
  \left( \phantom{\sum_{\bf x}}\hskip -0.22in
    {\cal A}_k^\dagger({\bf  y_m},0) \right)^{n_{km}}  
  \Bigg\rangle
  \ ,
  \label{eq:mk}
\end{eqnarray}
where ${\cal N}_i = \sum_j\ n_{ij}$ is the total number of mesons of
species $i$, and the subscript in $C_{\bf n} (t)$ labels the number of
each species from each source,
\begin{eqnarray} {\bf n} & = &\left(
    \begin{array}{cccc}
      n_{11}&n_{12}&...&n_{1m}\\
      \vdots &\vdots &\vdots &\vdots \\
      n_{k1} & n_{k2}&...&n_{km}
    \end{array}
  \right)
  \ \ \ .
  \label{eq:nvecdef}
\end{eqnarray}
The ${\cal A}_i({\bf y},t)$ are defined immediately after
eq.~(\ref{eq:mm}).  It is straightforward to show that
\begin{eqnarray}
  && C_{\bf n}(t) \ = \  
  \left(\ \prod_i\ {\cal N}_i!\ \right)\ 
  \left\langle\ \prod_{i,j} 
  \left(\ \overline{\eta}\ P_{ij}\ \eta\ \right)^{n_{ij}}\ \right\rangle
  \ \ \ ,
  \label{eq:mkGRASS}
\end{eqnarray}
where the $\eta$ are $m\times N$-component Grassmann variables, and
the $P_{ij}$ are $\overline{N}\times\overline{N}$ dimensional
matrices, where $\overline{N}=m\times N$, which are generalizations of
the $P_j$ defined in eq.~(\ref{eq:GRASSMANNmpi}) with an additional
species index, $i$. They are defined as
\begin{eqnarray}
  P_{ij} & = & 
  \left(
    \begin{array}{c|c|c|c}
      0&0&...&0 \\
      \hline
      \vdots & \vdots & ... & \vdots \\
      \hline
      \left(A_i\right)_{j1}(t)&\left(A_i\right)_{j2}(t)&\ \  ... \ \ & \left(A_i\right)_{jm}(t) \\
      \hline
      0&0& ... & 0\\
      \hline
      \vdots & \vdots & ... & \vdots \\
      \hline
      0&0& ... & 0
    \end{array}
  \right)
\end{eqnarray}
where the 
\begin{eqnarray}
  \left(\ A_i\ \right)_{ab} & = & 
  \sum_{\bf x}\ S({\bf x},t;{\bf y}_b,0)\ S_i^\dagger ({\bf x},t;{\bf
    y}_a,0)\ 
  \ \ \ ,
  \label{eq:Sijdef}
\end{eqnarray}
are $N\times N$ dimensional matrices, one for each flavor, $i$, and
pair of source indices, $a$ and $b$.  These correlators can be
expressed as
\begin{eqnarray}
  && C_{\bf n}(t) \ =\  
  (-)^{\overline{\cal N}}\ 
  {\prod_i \ {\cal N}_i! \over (\overline{N}-\overline{\cal N})!}\ 
  \epsilon^{a_1 ...a_{\overline{N}-\overline{\cal N}}
    \alpha_1  ... \alpha_{n_{11}}\alpha_{n_{11}+1}...
    \alpha_{\overline{\cal N}} } \ 
  \epsilon_{a_1 ...a_{\overline{N}-\overline{\cal N}}
    \beta_1   ... \beta_{n_{11}}
    \beta_{n_{11}+1}... \beta_{\overline{\cal N}} }  \ 
  \nonumber\\
  && 
  \qquad  \qquad  \qquad  \qquad   
  \left[ \ P_{11}(t)\ \right]_{\alpha_1}^{\beta_1}  ...
  \left[ \ P_{11}(t)\ \right]_{\alpha_{n_{11}}}^{\beta_{n_{11}}} 
  ...
  \left[ \ P_{km}(t)\ \right]_{\alpha_{\overline{{\cal N}}-n_{k,m}
      +1 }}^{\beta_{\overline{{\cal N}}-n_{k,m} +1 }} 
  \left[ \ P_{km}(t)\ \right]_{\alpha_{\overline{{\cal N}}}}^{\beta_{\overline{{\cal
          N}}} }
  \nonumber\\
  & = & 
  (-)^{\overline{\cal N}}\ 
  {\left(\ \prod_i\ {\cal N}_i!\right)\  \ \left(\ \prod_{i,j}\ n_{ij}!
    \right)\ 
    \over \overline{\cal N}!}
  \ 
  \langle\ T_{  {\bf n } }\ \rangle
  \ \ \ \ ,
  \label{eq:mkeps}
\end{eqnarray}
where $\overline{\cal N} = \sum_i\ {\cal N}_i$ is the total number of
mesons in the system, with $\overline{\cal N} \le \overline{N}$.  The
$T_{{\bf n} }$ satisfy the recursion relation
\begin{eqnarray}
  T_{  {\bf n}+{\bf 1}_{rs} }
  & = & 
  \sum_{i=1}^k\ 
  \sum_{j=1}^m\ 
  \ 
  \langle\ T_{ {\bf n}  + {\bf 1}_{rs} - {\bf 1}_{ij}  }\ \rangle\ P_{ij}
  \ -\
  \overline{\cal N}\  T_{  {\bf n}  + {\bf 1}_{rs} - {\bf 1}_{ij} }\ \ P_{ij}
  \ \ \ \ ,
  \label{eq:genRECUR}
\end{eqnarray}
where we have introduced the notation
\begin{eqnarray} {\bf 1}_{ij} \ & = &\left(
    \begin{array}{cccc}
      0 & 0&\cdots& 0\\
      \vdots &\vdots &\hdots\  1\ \hdots&\vdots \\
      0 & 0&\cdots&0
    \end{array}
  \right)
  \ \ \ .
  \label{eq:nplusvecdef}
\end{eqnarray}
where the non-zero value is in the $(i,j)^{\rm th}$ entry.
Defining ${\cal U}_j = \sum_i \ n_{ij}$ to be the number of mesons
from the $j^{\rm th}$ source, it is clear that the correlation
function vanishes when ${\cal U}_j>N$ for any source $j$.
Eq.~(\ref{eq:genRECUR}) is the main result of this work. It allows the
correlation functions of essentially arbitrary meson systems to be
evaluated. The correlation function defined in eq.~(\ref{eq:mk}) can
accommodate a total of $12L^3$ mesons where $L$ is the 
number of lattice sites in each spatial direction
of
the lattice. This filled system would correspond to a total meson
density of $1/b^3$ where $b$ is the lattice spacing. To go to even
higher densities, sources (and sinks) must be placed on multiple
time-slices.

\section{Discussion}
\label{sec:discussion}

\noindent
In this work, we have developed recursion relations that enable the
calculation of the correlation function of a system composed of
arbitrary numbers of mesons of different species generated from quark
propagators originating from different sources.  These recursion
relations will allow for Lattice QCD calculations of many-body systems
that will elucidate the phase transitions (or cross-overs) that are
expected to exist in QCD at finite meson density \cite{Son:2000xc,
  Kogut:2004zg, Sinclair:2006zm, deForcrand:2007uz}, and will also
allow for the exploration of systems at high density.  Further, they
will allow for fixed-density calculations in multiple lattice volumes,
thereby providing a means to control the finite-volume systematic
effects of such calculations.

The recursion relations scale to very large meson number and enable
the calculation of the correlation functions of systems composed of
(more precisely, with the quantum numbers of) large numbers of mesons
of different species, which are presently not practical to evaluate.
A further advantage of the recursive construction is that it
significantly reduces the overall computational cost. Each application
of the recursion requires only a single matrix multiplication for each
type of meson (or source) involved and a few additional scalar
operations. The memory requirements are also modest.  In contrast, the
expressions for the fully evaluated contractions (as displayed in
Ref.~\cite{Detmold:2008fn} for the single species, single source case)
contain a number of terms that grows exponentially in the number of
mesons, with an exponent that rapidly increases with the complexity of
the system (number of sources or species of meson). For the single
source, single species case, the two methods are comparable, but for
more complicated systems, the recursive approach requires fewer
operations to evaluate the corresponding correlation functions.

The recursive method can also be applied to other types of meson
systems such as those involving annihilation type diagrams (for
example, multiple $\pi^0$ systems), However, the construction of the
equivalent of the $A_{ij}(t)$ objects defined above is
computationally expensive.  In the case of baryons, or mixed
meson-baryon systems, recursive relations exist, but are much more
difficult to generalize.  This is currently under investigation.

\section{Acknowledgments}
\noindent
We would like to thank the NPLQCD collaboration for useful discussions
during this project. The work of MJS was supported in part by the
U.S.~Dept.~of Energy (DOE) under Grant No.~DE-FG03-97ER4014.  The work
of WD is supported in part by Jefferson Science Associates, LLC under
U.S. Dept. of Energy contract No. DE-AC05-06OR-23177 and by DOE
Outstanding Junior Investigator Award DE-SC000-1784 and by the
Thomas F. Jeffress and Kate Miller Jeffress Memorial Trust. The
U.S. government retains a non-exclusive, paid-up irrevocable,
world-wide license to publish or reproduce this manuscript for
U.S. government purposes.

%
%

\end{document}